\documentclass[12pt]{article}
\pdfoutput=1
\usepackage[letterpaper, margin=1in]{geometry}
\usepackage{amsmath, amssymb}
\usepackage{cite}
\usepackage{graphicx}
\usepackage[font=small,labelfont=bf]{caption}

\usepackage{xcolor}
\usepackage{slashed}
\usepackage{cancel}
\usepackage{bm}
\usepackage{verbatim}
\usepackage{tabularx}

\definecolor{DarkRed}{rgb}{0.6,0,0}
\definecolor{DarkGreen}{rgb}{0,0.6,0}
\definecolor{DarkBlue}{rgb}{0,0,0.6}

\usepackage{setspace}
\usepackage{authblk}
\usepackage{physics}
\usepackage{caption}
\usepackage{subcaption}

\usepackage{bbm}

\usepackage{dsfont}

\usepackage{etoolbox}
\appto\appendix{\addtocontents{toc}{\protect\setcounter{tocdepth}{1}}}

\let\OLDthebibliography\thebibliography
\renewcommand\thebibliography[1]{
	\OLDthebibliography{#1}
	\setlength{\parskip}{3.0pt plus 2.5pt minus 1.0pt}
	\setlength{\itemsep}{3.0pt plus 2.5pt minus 1.0pt}
}

\usepackage[colorlinks=true,linktocpage=true,linkcolor=DarkBlue,citecolor=DarkGreen,urlcolor=DarkGreen]{hyperref}

\usepackage[greek,english]{babel}

\newcommand{\be}{\begin{equation}}
\newcommand{\ee}{\end{equation}}
\def\ba{\begin{aligned}}
\def\ea{\end{aligned}}
\def\bga{\begin{aligned}}
\def\eda{\end{aligned}}
\def\bp{\begin{pmatrix}}
\def\ep{\end{pmatrix}}
\def\bgp{\begin{pmatrix}}
\def\edp{\end{pmatrix}}

\def\hc{\textrm{h.c.}}

\DeclareMathOperator{\diag}{diag}

\newcommand{\thetabar}{\overline{\theta}}

\newcommand{\Mpl}{M_{\textrm{Pl}}}
\newcommand{\LCP}{\Lambda_{\textrm{CP}}}
\newcommand{\TRH}{T_{\rm reh}}
\newcommand{\Hinf}{H_{\mathrm{inf}}}

\newcommand{\Sec}[1]{Section~\ref{#1}}

\newcommand{\Subsec}[1]{Subsec.~\ref{#1}}

\begin{document}
 
\title{\bf 
\Large Chiral Nelson--Barr Models: Quality and Cosmology
}
\author{Pouya Asadi$^{1}$,~
Samuel Homiller$^{2}$,~ 
Qianshu Lu$^{2}$,~ and 
Matthew Reece$^{2}$\\\vskip-0.75em
{\small \color{gray} \texttt{pasadi (@uoregon.edu), shomiller, qianshu$\_$lu, mreece (@g.harvard.edu)}} \\\vskip0.25em
\small \textit{${}^1$Institute for Fundamental Science and Department of Physics,} \\ \textit{University of Oregon, Eugene, OR 97403, USA} \\\vskip0.35em
\small \textit{${}^2$Department of Physics, Harvard University, Cambridge, MA 02138, USA}
}

\date{\today}

\maketitle

\begin{abstract}
\begin{spacing}{1.05}\noindent
\normalsize
It was recently shown that domain walls from the spontaneous breaking of CP symmetry are exactly stable, and must be inflated away to recover a viable cosmology. We investigate the phenomenological implications of this result in Nelson--Barr solutions of the strong CP problem. Combined with the upper bound on the scale of spontaneous CP breaking necessary to suppress contributions from dangerous, nonrenormalizable operators to $\bar{\theta}$, this puts an upper bound on the scale of inflation and the reheating temperature after inflation. Minimal Nelson--Barr models are therefore in tension with thermal leptogenesis, models of large-field inflation, or potential future observations of signals from topological remnants of an unrelated, subsequent phase transition. We study how extending Nelson--Barr models with a new, continuous chiral gauge symmetry can ameliorate this tension by forbidding the dangerous dimension-five operators. In particular, we show that gauging a linear combination of baryon number and hypercharge allows for an economic, anomaly-free extension of the minimal Nelson--Barr model, and discuss the phenomenological implications.
\end{spacing}
\end{abstract}

\begin{spacing}{1.15}

%\newpage 

\tableofcontents

\vskip 1cm

%%%%%%%%%%%%%%%%%%%%%%%%%%%%%%%%%%%%%%%%%%%%%%%%%%%%%%%%%%

%%%%%%%%%%%%%%%%%%%%%%%%%%%%%%
%%%%%%%%%%%%%%%%%%%%%%%%%%%%%%
\section{Introduction}

A significant milestone in our understanding of Quantum Field Theory was the realization that
the $\theta$-term in Yang-Mills theory,
%boe%
\be
\frac{\theta}{16\pi^2} \int \dd^4x\, \tr(G_{\mu\nu}\widetilde{G}^{\mu\nu}),
\ee
%eoe%
has physical consequences that depend on the coefficient $\theta \in [0, 2\pi)$~\cite{Belavin:1975fg, tHooft:1976rip, Jackiw:1976pf, Callan:1976je}. 
In the presence of fermions, this topological term is related to the chiral anomaly, and the invariant angle is
%boe%
\be
\thetabar = \theta + \arg\det(m_u) + \arg\det(m_d) ,
\label{eq:defth}
\ee
%eoe%
where $\theta$ is the ``bare'' Yang-Mills parameter and $m_u$, $m_d$ are the mass matrices of the up and down-type quarks, respectively.
The QCD Lagrangian violates CP for $\thetabar \neq 0, \pi$.  

When matched onto the chiral Lagrangian at low energies, $\thetabar$ parameterizes effects such as CP-violating pion-nucleon couplings and an electric dipole moment (EDM) for the neutron. 
Experimentally, however, the neutron EDM is constrained to be $\lesssim 10^{-26}\,e\,\textrm{cm}$ \cite{Abel:2020pzs,Workman:2022ynf}, which translates to 
%boe%
\be
\thetabar \lesssim 10^{-10} .
\label{eq:boundth}
\ee
%eoe%
This is in sharp contrast to the observed CP violating phase in the CKM matrix of the Standard Model (SM), which is $\mathcal{O}(1)$. This apparent fine-tuning is known as the strong CP problem.

The topological nature of the $\theta$ term in the QCD Lagrangian makes the strong CP problem particularly interesting from a UV perspective, despite being an IR observable. 
Moreover, there are essentially two well-known, viable\footnote{From Eq.~\eqref{eq:defth} one can see that if any of the SM quarks were massless, $\thetabar$ could be shifted away to zero, solving the strong CP problem. Lattice studies have firmly established nonzero masses for the SM quarks~\cite{Workman:2022ynf}, ruling out this solution.} classes of solutions to the strong CP problem: 
the existence of an anomalous $U(1)_{\textrm{PQ}}$ symmetry~\cite{Peccei:1977hh, Peccei:1977ur}, or the spontaneous breaking of an exact spacetime symmetry, such as parity~\cite{Mohapatra:1978fy,Babu:1988mw,Babu:1989rb,Barr:1991qx} or CP~\cite{Nelson:1983zb,Nelson:1984hg,Barr:1984qx,Barr:1984fh}. 
While each of these solutions treats CP on a very different footing, both require that the SM be extended by a new, nearly exact symmetry.
The strong CP problem is thus a particularly interesting testing ground for understanding physics at the highest energy scales.

Such setups can naturally give rise to many interesting dynamics and observables in the early universe. While there has been a recent resurgence in study of such effects in $U(1)_{\mathrm{PQ}}$ solutions (for recent reviews, see Ref.~\cite{Hook:2018dlk} and references therein), the parity and CP solutions have only begun to receive their due attention. 

Unlike CP violation in the SM, which is accompanied by suppressed flavor mixing or neutrino masses, 
the SM violates parity maximally. Therefore parity solutions to the strong CP problem, which need to generate a large parity violation in the SM while keeping $\bar{\theta}$ suppressed, require a significant amount of additional structure. For recent work on the parity solution, see, e.g., Refs.~\cite{Chakdar:2013tca,Hall:2018let,Dunsky:2019api,Craig:2020bnv,deVries:2021pzl}.

In this paper we will focus on solutions in which CP is assumed to be an exact symmetry of nature, spontaneously broken at a high scale. 
This solution has many guises---see, e.g.,~\cite{Hiller:2001qg,Hiller:2002um, Branco:2003rt,Vecchi:2014hpa,Dine:2015jga,Schwichtenberg:2018aqc,Cherchiglia:2019gll,Perez:2020dbw,Valenti:2021rdu,Valenti:2021xjp,Fujikura:2022sot,Girmohanta:2022giy} for a collection of classic and recent papers---but the general setup is referred to as the Nelson--Barr solution.
In a UV-complete, holographic quantum gravity theory, if CP is to be a good symmetry at high scales it must be a gauge symmetry and not a global one~\cite{Harlow:2018tng}. This is an instance of the more general absence of exact global symmetries in quantum gravity~\cite{Zeldovich:1976vq,Banks:1988yz,Banks:2010zn,Harlow:2018tng}. It  was recently shown in Ref.~\cite{McNamara:2022lrw} that when a gauged, orientation-reversing spacetime symmetry like CP is spontaneously broken, the associated domain walls are exactly stable because they carry a type of gauge charge. This fact about quantum gravity has dramatic, physical consequences for the cosmology of spontaneous CP violation. 

In the absence of a mechanism to destroy them, these domain walls must have their energy density diluted via inflation, implying that inflation must occur {\em after} the spontaneous breaking of CP. 
As we will discuss in detail, this condition is in opposition to the requirement that the scale of CP breaking should be small enough to maintain $\thetabar \simeq 0$ to good approximation, i.e., the Nelson--Barr \textit{quality problem}. 
This implies an upper bound on the scale of inflation, which in turn imposes an upper bound on the subsequent tensor-to-scalar ratio. 
We will discuss how many of the simplest models of inflation are incompatible with this setup. 
We also study the gravitational wave signals from topological defects that arise from breaking of unspecified internal symmetries after such a low reheating temperature; we find that detection of gravitational waves from a string network rules out minimal Nelson--Barr constructions. 
We also investigate the compatibility of such constructions with a few popular baryogenesis mechanisms; in particular, we will show that minimal Nelson--Barr constructions are at odds with thermal leptogenesis. 

Finally, we will show that this tension can be alleviated in simple modifications of the Nelson--Barr constructions where the new heavy quarks transform chirally under a new continuous gauge symmetry. 
This solves the quality problem by postponing dangerous operators to dimension six, and allows inflation to happen at higher scales. Such chiral Nelson-Barr models  were previously studied, in a composite setting, in Ref.~\cite{Valenti:2021xjp}. Compositeness leads to a  number of predictions about mass scales that are quite different from the simple elementary model that we propose. 
We study the implications of the higher inflation scale for models of inflation, gravitational wave signals, and the thermal leptogenesis setups.

\medskip

In the next section, we review how an ultraviolet CP symmetry can solve the strong CP problem, and recap the arguments of Ref.~\cite{McNamara:2022lrw}. 
In Sec.~\ref{sec:nb_intro} we review the simplest models with spontaneous CP breaking and show that the scale at which CP is broken cannot be arbitrarily large while keeping $\thetabar$ adequately small---the Nelson--Barr ``quality problem''. 
Sec.~\ref{sec:cosmology} is devoted to a detailed discussion of the consequences of these arguments in cosmology, particularly as it pertains to models of inflation and attempts to dynamically generate the baryon asymmetry. 
In Sec.~\ref{sec:chiral_model}, we discuss a model in which the bound on the scale of CP breaking is alleviated and baryon asymmetry can be generated via thermal leptogenesis. 
We conclude in Sec.~\ref{sec:conclusions} and discuss interesting directions for future work.

%%%%%%%%%%%%%%%%%%%%%%%%%%%%%%
%%%%%%%%%%%%%%%%%%%%%%%%%%%%%%
\section{CP as a Gauge Symmetry}
\label{sec:cpgauge}

While CP is not a symmetry of the Standard Model~\cite{Landau:1957tp, Lee:1957qq, Christenson:1964fg}, it is noteworthy that CP-violating effects are only observed along with flavor violation in the quark mass-mixing matrix. 
As a consequence, the running of the $\thetabar$ parameter appears only at seven loops in the SM~\cite{Ellis:1978hq}, so that if $\thetabar$ is fixed to zero at a high scale, it remains small in the IR. (While the running of $\thetabar$ appears at seven loops, Khriplovich found that there is a finite contribution at four loops from the ``Cheburashka diagram''~\cite{Khriplovich:1985jr}.)
These facts, along with the strong bound on $\thetabar$ from neutron EDM~\cite{Workman:2022ynf} experiments, lend credence to the idea that CP might be an exact (or extremely good) symmetry of nature which is spontaneously broken at high scales. 

If CP is assumed to be an exact symmetry, there exists a basis in which all the Lagrangian parameters are real and $\thetabar = 0$ in the UV.
The challenge in constructing viable models is to generate a CKM phase that is not suppressed by factors of $v / \LCP$, where $v = 246\,\textrm{GeV}$ is the weak scale and $\LCP$ is the scale at which CP is spontaneously broken, while still maintaining $\thetabar \lesssim 10^{-10}$.

Models which do this successfully, first developed by Nelson~\cite{Nelson:1983zb,Nelson:1984hg} and generalized by Barr~\cite{Barr:1984qx, Barr:1984fh}, are known as the Nelson--Barr solution to the strong CP problem. 
(Earlier proposals relying on spontaneous CP breaking could not accommodate the large CP violating angle in the CKM matrix without generating large $\bar{\theta}$ \cite{Beg:1978mt, Georgi:1978xz, Segre:1979dx,Goffin:1979mt,Barr:1979as,Barr:1980mq,Masiero:1981zd,Barr:1983rz}.)
The idea of spontaneous CP violation can be naturally embedded in more sophisticated models.
Notable examples include Hiller-Schmaltz models~\cite{Hiller:2001qg, Hiller:2002um}, which use supersymmetry to protect against corrections to $\thetabar$ via nonrenormalization theorems, and models of twisted split fermions, which embed a similar idea in extra dimensions~\cite{Grossman:2004rm, Harnik:2004su}. Motivated by the connection to the CKM phase and flavor-violation, Nir and Rattazzi constructed a model of flavor~\cite{Nir:1996am} in which CP is spontaneously broken by the flavon fields of a Froggatt-Nielsen setup \cite{Froggatt:1978nt}. 
This idea was recently extended to the lepton sector in Refs.~\cite{Aloni:2021wzk, Nakai:2021mha}.

The common thread in all of these proposals is that CP is assumed to be an {\em exact} symmetry of nature. If CP is not an exact symmetry, there is no principled reason to believe it is not violated by renormalizable operators in the theory, in addition to nonrenormalizable ones.  Indeed, in a theory of quantum gravity, it is well understood that global symmetries are badly broken in the UV~\cite{Nomura:2019qps, Cordova:2022rer}, and any good symmetry observed at low energies must be an emergent consequence of gauge symmetries and locality in the underlying theory. This reasoning applies to discrete spacetime symmetries as well as internal symmetries. 
From this perspective, it does not make sense to consider models in which CP is an approximate symmetry that is broken only through higher-dimension Planck-suppressed operators.

It is therefore vital to consider the physical consequences of demanding CP to be an exact (gauge) symmetry in the UV. This possibility was first considered in Refs.~\cite{Strominger:1985it, Dine:1992ya, Choi:1992xp}, but has otherwise received relatively little attention until recently~\cite{McNamara:2022lrw}.
As with any discrete symmetry, the spontaneous breaking of CP gives rise to domain walls separating different vacua. 
Assuming that the early universe went through a phase transition to the spontaneously broken CP phase, a network of such domain walls will form~\cite{Zeldovich:1974uw, Kibble:1976sj, Zurek:1985qw}. The energy density in a domain wall network scales as the inverse of the scale factor, and they therefore quickly dominate the energy density of the universe. If these domain walls are stable, they therefore pose a serious problem for cosmology. (For pedagogical reviews on this topic, see Refs.~\cite{Kibble:1980mv,Vilenkin:1984ib,Vilenkin:2000jqa}.)

The key result of Ref.~\cite{McNamara:2022lrw} is that if CP is a gauge symmetry, there is no local dynamical process by which CP domain walls can be destroyed, i.e., they are exactly stable. This contrasts with discrete internal gauge symmetries, for which the corresponding domain wall networks can potentially be destroyed by cosmic strings~\cite{Vilenkin:1982ks, Kibble:1982dd, Kibble:1982ae, Lazarides:1982tw, Everett:1982nm} (see also Ref.~\cite{Kim:1986ax} for a review of these ideas). Ref.~\cite{McNamara:2022lrw} demonstrated that the analogous process is topologically forbidden in the case of orientation-reversing spacetime symmetries like CP.

The only way to reconcile a spontaneously broken, exact CP symmetry with cosmology is if the energy density carried by the CP domain wall network is extremely suppressed. This can be achieved if inflation occurs {\em after} the spontaneous breaking of CP. 
Combined with the fact that the scale of spontaneous CP breaking cannot be arbitrarily high (see Sec.~\ref{subsec:quality_problem}), this puts an upper bound on the the scale of inflation ($\Hinf$) and the reheating temperature ($\TRH$) after  inflation. 
This imposes a nontrivial constraint on inflation and the subsequent cosmological evolution of the universe; we will discuss these consequences in Sec.~\ref{sec:cosmology}.

%%%%%%%%%%%%%%%%%%%%%%%%%%%%%%
%%%%%%%%%%%%%%%%%%%%%%%%%%%%%%
\section{Nelson--Barr Models and the Scale of Spontaneous CP Breaking}
\label{sec:nb_intro}

In this section we review the basics of Nelson--Barr solutions to the strong CP problem. 
We will mostly focus on a nearly minimal model, which illustrates the basic ideas and serves as a basis for the extension introduced in Sec.~\ref{sec:chiral_model}. 
As we will see in Sec.~\ref{subsec:quality_problem}, the scale at which CP is spontaneously broken cannot be arbitrarily large, without spoiling the solution to the strong CP problem.

%%%%%%%%%%%%%%%%%%%%%%%%%%%%%%
\subsection{A Minimal Model}
\label{subsec:bbp}

We consider a modest extension to the ``minimal'' model proposed by Bento, Branco and Parada (BBP)~\cite{Bento:1991ez}. 
We introduce a set of vectorlike, $SU(2)_L$ singlet quarks $\bar{D}$ and $D$ with charge $\pm 1/3$, as well as a set of $N$ complex scalars $\eta_a$ that are singlets under the SM.\footnote{The original BBP model used only one complex scalar, with a $Z_2$ symmetry. The straightforward extension to $N$ scalars was discussed in, e.g., Ref.~\cite{Dine:2015jga}.}
We further impose a $Z_N$ gauge symmetry under which
%boe%
\be
\eta_a \to \mathrm{e}^{2\pi i k / N} \eta_a, \quad
D \to \mathrm{e}^{-2\pi i k / N} D, \quad
\bar{D} \to \mathrm{e}^{2\pi i k / N} \bar{D},
\ee
%eoe%
with the SM fields neutral. The allowed, renormalizable mass and Yukawa couplings for the down-type quarks are therefore given by
%boe%
\be
\mathcal{L} \supset \mu_D D \bar{D} + (\lambda_d)^i{}_j Q_i H^c \bar{d}^j - f^a_i \eta_a D \bar{d}^i + \hc
\label{eq:NBL}
\ee
%eoe%
where $Q$ and $\bar{d}$ refer to left-handed (LH) and right-handed (RH) SM quarks, $H$ is the Higgs field, $\lambda_d$ and $f$ are Yukawa couplings, $i, j = 1, 2, 3$ are flavor indices, and $a = 1, \dots N_\eta$ labels the complex scalars. 
Assuming CP invariance, we can take $\mu_D$ and all the components of $f$ and $\lambda_d$ to be real. We will further assume (without loss of generality) that we are in the flavor basis in which the up quark mass matrix is real diagonal.

The scalar sector of the Lagrangian includes
%boe%
\be
\mathcal{L} \supset \lambda_{ab} \eta_a \eta_b^{\dagger}|H|^2 + \gamma_{abcd} \eta_a \eta_b \eta_c^{\dagger}\eta_d^{\dagger} .
\ee
%eoe%
For general parameters in the potential, some subset of the $\eta_a$ can acquire vacuum expectation values (vevs) that will in general be complex.\footnote{See Ref.~\cite{Haber:2012np} for a detailed discussion on the conditions for spontaneous CP violation.} We write
%boe%
\be
\sum_a f_i^a \langle \eta_a \rangle \equiv F_i
\ee
%eoe%
where $i$ is a down-type flavor index and $\langle \eta_a \rangle \sim \LCP\,e^{i\alpha}$, 
where $\LCP$ is the scale of spontaneous CP breaking and $\alpha$ is a phase, assumed to be order one.

At tree level, the mass matrix for the down-type quarks in Eq.~\eqref{eq:NBL} can be written in block form,
%boe%
\be
\label{eq:MD_tree}
\mathcal{L} \supset \bp Q & D\ep  \mathcal{M}_0 \bp \bar{d} \\ \bar{D} \ep,
\qquad
\mathcal{M}_0 = \bp m_d & 0 \\ F & \mu_D \ep,
\ee
%eoe%
where $(m_d)^i{}_j = (\lambda_d)^i{}_j v / \sqrt{2}$ and $F$ should be viewed as a row vector with entries $F_i$. 
Note that because the upper-right block is zero and all the entries of $m_d$ and $\mu_D$ are real, $\det \mathcal{M}_0$ is real. Since CP is a symmetry of the theory, the bare $\theta$ parameter vanishes, so $\thetabar = 0$ in the renormalizable theory at tree level.

The mass matrix $\mathcal{M}_0$ can be diagonalized with the usual bi-unitary transformation, $U_L^{\dagger} \mathcal{M}_0 U_R = \diag(\bar{m}, m_D)$, where $\bar{m}$ is the diagonal matrix of real, SM quark masses and $m_D$ the physical mass of the vectorlike quarks. For heavy new $D$ quarks, the upper-left $3 \times 3$ block of $U_L$ can be identified as the usual CKM matrix, $V$. Noting that $U_L^{\dagger} \mathcal{M}_0^{\dagger} \mathcal{M}_0 U_L = \diag(\bar{m}^2, m_D^2)$, we find 
%boe%
\be
(V \overline{m}^2 V^{\dagger})^i_j \simeq (m_d m_d^T)^i_j + \frac{(m_d)^i_k F^{\dagger}_k F_l (m_d^T)^l_m}{F_p F^{\dagger}_p + \mu_D^2} \bigg( \delta^m_j + \frac{(\overline{m}^2)^m_j}{F_qF^{\dagger}_q + \mu_D^2}\bigg)
\equiv (m_0^2)^i_j + \mathcal{O}(v^2 / \LCP^2)
\ee
%eoe%
where we have defined the effective $3\times 3$ mass mixing matrix for the SM quarks,
%boe%
\be
(m_0^2)^i_j = (m_d)^i_k \Big( \delta^k_l + \frac{F^{\dagger}_k F_l}{F_p F^{\dagger}_p + \mu_D^2}\Big) (m_d^T)^l_j .
\label{eq:mCKM}
\ee
%eoe%
This is the desired result: the effective mixing matrix is approximately diagonalized by the CKM matrix, $V$. We see that while the matrix $m_d$ is real, the effective mixing matrix is generally complex, due to the complex vevs in $F_i$, with no ratios of $v / \LCP$. 
We want this diagonalizing matrix to have an $\mathcal{O}(1)$ phase, which requires $F^{\dagger}_p F_p \gtrsim \mu_D^2$, so that the second term in parentheses has entries of order unity. Note also that for $V$ to be approximately unitary, we must assume $F, \mu_D \gg m_d$.

%%%%%%%%%%%%%%%%%%%%%%%%%%%%%%
\subsection{Radiative Corrections to $\thetabar$}
\label{subsec:loop_corrections}

We now turn to radiative corrections to $\thetabar$ which, in the models we consider, can all be packaged as corrections to the down-quark mass matrix $\mathcal{M}_D$. We are interested in the quantity
%boe%
\be
\Delta\thetabar = \arg\det \mathcal{M}_D.
\ee
%eoe%
We expand $\mathcal{M}_D = \mathcal{M}_0 + \mathcal{M}_1$, where $\mathcal{M}_0$ is the tree-level mass matrix (Eq.~\eqref{eq:MD_tree}), and $\mathcal{M}_1$ characterizes one-loop corrections that are assumed to be small (so that the entries of $\mathcal{M}_0^{-1} \mathcal{M}_1$ are $ \ll 1$). 
We then write,
\be
\det \mathcal{M}_D = \det \mathcal{M}_0 \det\big(\mathbbm{1} + \mathcal{M}_0^{-1} \mathcal{M}_1\big) \simeq \det \mathcal{M}_0 \Big( 1 + \tr(\mathcal{M}_0^{-1} \mathcal{M}_1)\Big) .
\ee
Thus,
%boe%
\be
\Delta\thetabar \simeq \arg\Big( 1 + \tr(\mathcal{M}_0^{-1} \mathcal{M}_1)\Big) \simeq \Im\Big( \tr(\mathcal{M}_0^{-1} \mathcal{M}_1)\Big) .
\ee
%eoe%
Taking $\mathcal{M}_1$ to have the block form
%boe%
\be\label{eq:deltaM}
\mathcal{M}_1 = \bp \mathcal{M}^{(1)}_{Q\bar{d}} & \mathcal{M}^{(1)}_{Q\bar{D}} \\[0.5em] 
\mathcal{M}^{(1)}_{D\bar{d}} & \mathcal{M}^{(1)}_{D\bar{D}}, \ep
\ee
%eoe%
we find
%boe%
\be
\label{eq:thetacorrection}
\Delta\thetabar \simeq \Im\!\Big( m_d^{-1} \mathcal{M}^{(1)}_{Q\bar{d}} - \frac{1}{\mu_D}\big( F m_d^{-1} \mathcal{M}^{(1)}_{Q\bar{D}} + \mathcal{M}^{(1)}_{D\bar{D}}\big) \Big) ,
\ee
%eoe%
where all the SM flavor indices are suppressed to avoid notation clutter. Note that corrections to the $\mathcal{M}^{(1)}_{D\bar{d}}$ mass term do not appear in this expression.

\begin{figure}[t]
\centering
\includegraphics[width=0.9\linewidth]{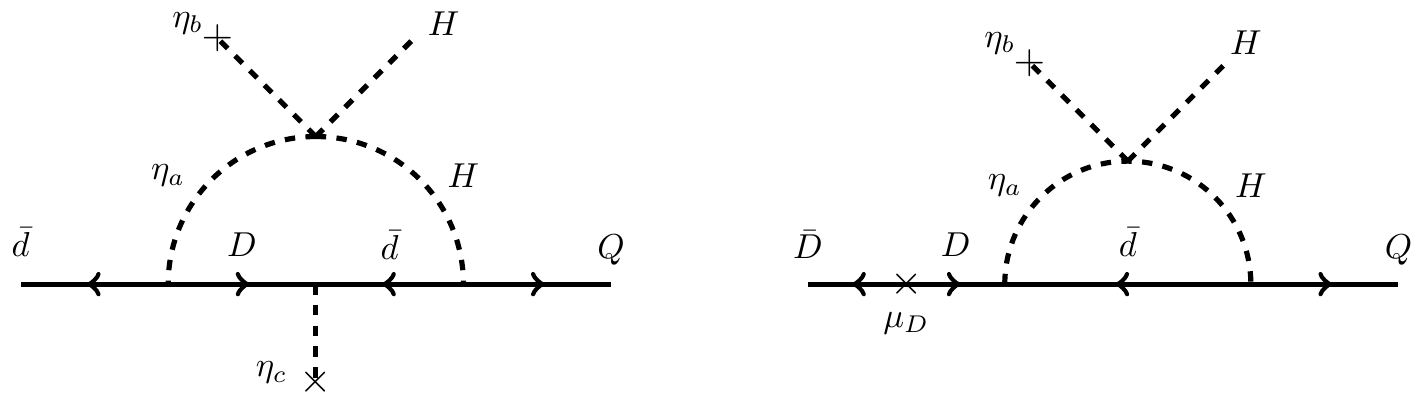}
\caption{Diagrams leading to corrections to $\Delta\thetabar$ at one loop.}
\label{fig:loop_diagrams}
\end{figure}

There are a number of possible diagrams that contribute to the various self-energies in $\mathcal{M}_1$ in this model, but the majority of them add up to give contributions that are manifestly real, and hence do not contribute to $\thetabar$. The exceptions are the classes of diagrams shown in Fig.~\ref{fig:loop_diagrams}, which contribute to $\mathcal{M}^{(1)}_{Q\bar{d}}$ and $\mathcal{M}^{(1)}_{Q\bar{D}}$, respectively. These can be evaluated straightforwardly in the small momentum limit, and we find, for instance,
%boe%
\be
i \left(\mathcal{M}^{(1)}_{Q\bar{d}} \right)_{ij}\sim \frac{i}{8\pi^2} \sum_{a, b, c} \frac{1}{m_{\eta}^2} \log\Big(\frac{v^2}{\LCP^2}\Big) (m_d)^k{}_i f_k^c\, f_j^{a} \lambda_{ab} \langle \eta_b \rangle \langle \eta_c^{\dagger} \rangle ,
\ee
%eoe%
where we have assumed all $\eta$ fields have comparable masses, denoted by $m_\eta$. 
Plugging into Eq.~\eqref{eq:deltaM}, this leads to a contribution to $\thetabar$ that is schematically,
%boe%
\be
\label{eq:delta_theta_loop}
\Delta\thetabar \sim \frac{1}{8\pi^2} f_k\, f_k\, \lambda \,\frac{\LCP^2}{m_{\eta}^2} \log\Big(\frac{v^2}{\LCP^2}\Big) .
\ee
%eoe%

%%%%%
%%%%%

The contribution from the second set of diagrams contributing to $\mathcal{M}^{(1)}_{Q\bar{D}}$ is essentially the same.\footnote{At two-loop order, new diagrams proportional to $\gamma_{abcd}$ correct the $\theta$ parameter~\cite{Nelson:1983zb,Dine:2015jga}. For $\gamma_{abcd} \sim \lambda_{ab}$, this contribution will be subdominant to the one-loop corrections above by a few orders of magnitude.} The sum of all these contributions must be $\lesssim 10^{-10}$ to preserve the solution to the strong CP problem. 
Absent any surprising cancellations, this requires the Yukawa couplings, $f$ and/or the quartic couplings $\lambda$ to be very small. 
As noted in Ref.~\cite{Bento:1991ez}, the quartic couplings $\lambda$ also shift the mass term for the SM Higgs doublet at tree level once the $\eta_a$ acquire a vacuum expectation value. This leads to a significant fine-tuning in the Higgs potential if the scale of spontaneous CP breaking is assumed to be far above the TeV scale. As in Ref.~\cite{Bento:1991ez}, we will assume that whatever physics is responsible for solving the electroweak hierarchy problem also suppresses these contributions,\footnote{In supersymmetry, for instance, these quartic couplings only arise from coupling to supersymmetry breaking, and are of order $|F|^2/\Mpl^4$, where $\sqrt{F}$ is the mass scale of supersymmetry breaking.} and that $\lambda$ is small enough so that the $f$ can be assumed to be $\mathcal{O}(1)$, with no other tuning necessary in the model. There is a one-loop correction to the quartic $\lambda$ of order
\begin{equation} \label{eq:lambdaloopcorrection}
\delta \lambda_{ab} \sim \frac{1}{16\pi^2} f^a_i f^b_j (\lambda_d)^k{}_i (\lambda_d)^k{}_j \log\frac{\Lambda_\mathrm{UV}^2}{\LCP^2}.
\end{equation}
Estimating the maximal contribution from the $b$-quark Yukawa, this combined with Eq.~\eqref{eq:delta_theta_loop} is consistent with $f \sim 10^{-1}$ and $\lambda_{ab} \sim 10^{-7}$ without tuning to reduce $\thetabar$.

The new, color-charged fermions $D$ and $\bar{D}$ must be at least heavy enough to evade direct searches at the LHC, implying $\mu_D \gtrsim \textrm{TeV}$. 
The condition that $\mu_D \lesssim f \LCP$ to obtain an $\mathcal{O}(1)$ CKM phase then requires (for $f \sim \mathcal{O}(1)$) that $\LCP \gtrsim 10^3~\textrm{GeV}$.

The new heavy quarks also contribute electric and chromoelectric dipole moments for the down-type quarks, which are stringently constrained by searches for the mercury and neutron EDMs~\cite{Baker:2006ts, Ellis:2008zy, Graner:2016ses}. In the minimal models we are considering, the nonvanishing contributions come from diagrams similar to Fig.~\ref{fig:loop_diagrams}, left, with an additional gluon or photon attached to the internal quark line. As a result, these contributions are proportional to the same combination of couplings that appears in Eq.~\eqref{eq:delta_theta_loop}. Assuming this combination is small enough to obtain $\thetabar \lesssim 10^{-10}$, the resulting bounds on $\LCP$ from the mercury and neutron EDMs are several orders of magnitude weaker than the direct search constraints from the LHC.

Parenthetically, we note that these Nelson--Barr constructions naively have all the ingredients for a viable model of electroweak baryogenesis~\cite{Cohen:1990py, Cohen:1990it, Nelson:1991ab, Cohen:1993nk}. This includes a new source of CP violation ($\langle \eta \rangle$) and a modification of the Higgs potential (via the new scalars) that could potentially give rise to a strong first order electroweak phase transition.
The arguments above, however, suggest that these new scalars appear at much higher scales, and would therefore have minimal effects on the order of the electroweak phase transition. 
Thus, a Nelson--Barr extension of SM alone is not enough for a viable electroweak baryogenesis.

%%%%%%%%%%%%%%%%%%%%%%%%%%%%%%
\subsection{The Nelson--Barr ``Quality Problem''}
\label{subsec:quality_problem}

Beyond the radiative corrections, Nelson--Barr models suffer from a potential ``quality problem,'' first discussed in Ref.~\cite{Choi:1992xp} and more recently emphasized by Ref.~\cite{Dine:2015jga}. In the setup discussed above, there are contributions from dimension-five operators
\begin{equation}
\mathcal{L } \supset 
\frac{h_{ab}}{\Lambda_{\textrm{EFT}}} \eta_a^{\dagger}\eta_b D \bar{D} 
+
\frac{{h'}_a^i}{\Lambda_{\textrm{EFT}}} \eta_a^{\dagger} Q_i H^c \bar{D} ,
\label{eq:dangerous5}
\end{equation}
where $h$ and $h'$ are dimensionless couplings and $\Lambda_{\textrm{EFT}}$ is the cutoff of the theory. A naive estimate is that these operators give contributions to $\thetabar \sim \LCP / \Lambda_{\textrm{EFT}}$. Even assuming the cutoff is as large as the Planck scale, $\Mpl$, this would require $\LCP \lesssim 10^8\,\textrm{GeV}$. A lower scale for the cutoff would make this bound even stronger. 

A more careful estimate of the contributions from these operators can change this naive constraint.
The first operator from Eq.~\eqref{eq:dangerous5} gives rise to a nonzero contribution to $\bar{\theta}$ of,
\be
\label{eq:dim5_contribution_1}
\Delta\thetabar \sim h_{ab} \frac{\langle \eta_a^{\dagger}\eta_b\rangle}{\mu_D \Lambda_{\textrm{EFT}}} .
\ee
If $ \mu_D \sim \sqrt{\langle \eta_a^{\dagger}\eta_b \rangle} \sim \LCP$, demanding that $\thetabar \lesssim 10^{-10}$ leads to the naive constraint of $\LCP \lesssim 10^8\,\textrm{GeV}$. Note that getting a nonzero phase in the CKM matrix requires $\mu_D \lesssim \LCP$ (assuming $f \sim 1$), which further suggests $\LCP$ could be at an even lower scale.

The second operator from Eq.~\eqref{eq:dangerous5} gives
\be
\Delta\thetabar \sim \frac{\langle \eta_a \rangle \langle \eta_b \rangle}{\mu_D \Lambda_{\textrm{EFT}}} \big( f_i^b (\lambda_d^{-1})^i{}_j h'^j_a\big), 
\ee
where $\lambda_d$ is the SM Yukawa couplings of the down-type quarks to the Higgs. This contribution depends on the particular pattern of entries in ${h'}_a^j$ and $f_i^a$, and can change substantially depending on the underlying model of flavor.
Treating the $f_i^a$ and ${h'}_a^j$ as spurions under the approximate SM flavor symmetry, we see that their product transforms in the same way as the down-type Yukawa matrix. In a minimal flavor violation scenario~\cite{Chivukula:1987py,DAmbrosio:2002vsn}, we therefore expect this product to be proportional to $(\lambda_d)^i_j$, and the resulting contribution is of the same order as the one in Eq.~\eqref{eq:dim5_contribution_1}.

It is also possible that the underlying model of flavor is more complex, and there may be additional flavor spurions with different patterns of couplings than in the SM Yukawas. 
For instance, in models with Spontaneous Flavor Violation---which are all in the Nelson--Barr class---new flavor spurions can exist at intermediate scales, without introducing dangerous contributions to flavor-changing processes~\cite{Egana-Ugrinovic:2018znw, Egana-Ugrinovic:2019dqu}. 
In the most extreme case, we can imagine that the $h^j_a$ are $\mathcal{O}(1)$. Then, the most problematic entry is the $11$ component of $\lambda_d^{-1}$, which is $\sim 1/y_d$, where $y_d \sim 10^{-5}$ is the down quark Yukawa. 
In this case, we would require $\LCP$ to be near the TeV scale in order to avoid dangerous contributions to $\thetabar$---a much more stringent constraint than in minimal flavor violating models.

We have seen that, absent a way of avoiding dangerous dimension-five operators, the scale at which CP is spontaneously broken in Nelson--Barr models cannot be arbitrarily large, and in some cases must be less than $10^8~\textrm{GeV}$. This bound has momentous consequences for cosmology, to which we now turn.

%%%%%%%%%%%%%%%%%%%%%%%%%%%%%%
%%%%%%%%%%%%%%%%%%%%%%%%%%%%%%
\section{Implications for Cosmology}
\label{sec:cosmology}

The stability of CP domain walls implies that the scale of inflation should be below the CP breaking scale, which itself is bounded from above thanks to the Nelson--Barr quality problem. 
As argued above, depending on details of the flavor texture, this bound could be as severe as $\LCP \lesssim 10^3$~GeV; we will, however, focus on the minimal flavor violating scenario, which suggests $\LCP \lesssim 10^8$~GeV. For completeness, we also remind the reader that existing bounds from LHC suggest $\LCP \gtrsim 10^3$~GeV, leaving a relatively narrow range of viable $\LCP$ for minimal Nelson--Barr models discussed above. 

Since Hubble throughout inflation ($\Hinf$) only changes mildly, we take the $ \Hinf \lesssim \LCP \lesssim 10^8\,\textrm{GeV}$ constraint to apply for the entire duration of inflation, including the time when CMB-scale curvature perturbations left the horizon and when inflation ended. 
This has a wide range of implications for cosmology, because it not only affects inflationary observables and model building, but also constrains the temperature that the thermal plasma could have had after inflation, i.e., the reheating temperature $T_{\rm reh}$.

For simplicity, we will assume single-field inflation in the following discussions. Reheating is complete when 1) the equation of state of the total fluid of the universe is radiation-like, or $w = 1/3$, and 2) the system reaches local thermal equilibrium. 
Generically, after a transient period of kination or inflaton oscillations in a steeper power-law potential, the inflaton settles into a quadratic minimum after the end of inflation. 
The inflaton fluid therefore has a matter-like equation of state after inflation, and reheating cannot be complete unless the matter-like inflaton energy density is completely converted into radiation-like species. One mechanism of reheating is through the perturbative decay of the inflaton \cite{Abbott:1982hn,Albrecht:1982mp,Dolgov:1982th}. 

In a perturbative decay, the energy of the daughter particles cannot be higher than the mass of the parent, so we expect the resulting thermal bath to have a temperature $T_{\rm reh} < m_{\phi}$. (This assumes the particles do not undergo rapid number-changing interactions to heat up; if they do, then a backreaction effect would tend to shut the process off~\cite{Kolb:2003ke}. This is not a strict no-go theorem, but a strong expectation.) 
Observations of the CMB indicate that the scalar curvature perturbation is nearly scale invariant, with $n_s = 0.9649 \pm 0.0042$. Since $n_s - 1 \supset \Mpl^2 V_{\phi\phi}/V$, near scale invariance of the curvature perturbation implies that the curvature scale of the inflaton potential is slightly smaller than Hubble. Assuming the potential behaves reasonably smoothly, we expect the same to be true for the inflaton mass around the minimum. Therefore, if reheating happened perturbatively,  even if the daughter particle bath thermalized quickly at the decay time, the resulting temperature is expected to be lower than Hubble at the end of inflation: $T_{\rm reh} \lesssim \Hinf \lesssim 10^8\,\textrm{GeV}$.

A potential loophole in the above argument is preheating: a period of resonant, nonperturbative particle production driven by the coherent oscillation of the inflaton after inflation, where daughter particles are produced with energies much higher than the inflaton mass \cite{Traschen:1990sw, Dolgov:1989us, Shtanov:1994ce, Kofman:1994rk, Boyanovsky:1995ud, Yoshimura:1995gc, Kaiser:1995fb, Kofman:1997yn}. However, the momentum distribution of the daughter particles peaks at the instability bands of their classical equations of motion, far away from a thermal distribution. 
In the limited number of cases where the system is simulated with sufficiently long time, thermalization is found to be a time-consuming process, resulting in a thermal plasma with energy density much lower than the energy density of the system at the end of inflation \cite{Micha:2002ey, Micha:2004bv}. 
It is doubtful that preheating will be able to result in a thermal plasma with a temperature higher than $T_{\rm reh} \lesssim \Hinf \lesssim 10^8\,\textrm{GeV}$.

In what follows, we will investigate the implications of this bound on models of inflation, gravitational waves signals from topological defects (from an unrelated phase transition below $\LCP$), and some mechanisms for baryogenesis. In particular, we find that minimal Nelson--Barr models:
\begin{itemize}
    \item require a very suppressed tensor-to-scalar ratio, incompatible with large-field inflation;
    \item could be compatible with detectable gravitational wave signals from networks of domain walls, but not from cosmic strings; and
    \item are incompatible with thermal leptogenesis, but could be compatible with a few other baryogenesis mechanisms, given further model-building.
\end{itemize}
These conclusions motivate building more sophisticated Nelson--Barr models where $\Lambda_{\mathrm{CP}}$ and, subsequently, the scale of inflation can be pushed higher. We will present such a model in Sec.~\ref{sec:chiral_model}.

%%%%%%%%%%%%%%%%%%%%%%%%%%%%%%
\subsection{Implications for Models of Inflation}
\label{subsec:inflation}

From the terrestrial side, the constraint $\Hinf \lesssim 10^8\,\textrm{GeV}$ is interesting because it implies new physics related to inflation is within a few orders of magnitude of collider scales. However, this upper bound can suppress other early universe observables. 
For single-field inflation, the tensor-to-scalar ratio is a direct measurement of Hubble during inflation,
\be
r = \frac{\mathcal{P}_{T}}{\mathcal{P}_{\zeta}} \approx 9.65 \times 10^7 \frac{\Hinf^2}{\Mpl^2},
\ee
where we have plugged in the Planck 2018 observation for the scalar power spectrum, $\mathcal{P}_{\zeta} = 2.0989\times 10^{-9}$\cite{Planck:2018jri}. The current CMB constraint on the tensor-to-scalar ratio of $r < 0.036$ by the BICEP/Keck experiment\cite{BICEP:2021xfz} translates to a constraint on Hubble of $\Hinf \lesssim 5 \times 10^{13}\,\textrm{GeV}$. The constraint $\Hinf \leq \LCP$ means $r$ is bounded from above by
\be
r \leq 1.7\times 10^{-13}\left(\frac{\LCP}{10^8\,\textrm{GeV}}\right)^2.
\label{eq:rbound}
\ee

Aside from the tremendous experimental challenges to observe such a small value of $r$, this also imposes a stringent constraint on inflationary model building. In single-field inflation, $r$ is related to the slow-roll parameter $\epsilon = \frac{\Mpl^2}{2}\left(\frac{V'}{V}\right)^2$ by a simple multiplicative factor, $r = 16 \epsilon$. The constraint $r \lesssim 10^{-13}$ therefore means $\epsilon \lesssim 10^{-14}$, an extremely flat inflaton potential during inflation. The small tensor-to-scalar ratio also limits the total inflaton field evolution during inflation through the Lyth bound~\cite{Lyth:1996im}, 
\be
\frac{\Delta \phi}{\Mpl} = \Delta N_{\rm CMB}\sqrt{\frac{r}{8}},
\ee
where $\Delta N_{\rm CMB}$ is the number of $e$-folds that the universe expanded between when the fluctuations of the CMB scale left the horizon and when inflation ended. With the conventional approximation of $\Delta N_{\rm CMB} \approx 50$, we see that the constraint $r \lesssim 10^{-13}$ implies a total inflation field evolution far below the Planck scale,
\be
\frac{\Delta \phi}{\Mpl} \lesssim 10^{-6},
\ee
i.e., inflation is driven by a small-field model.

A small $\epsilon$ and small inflaton field excursion to satisfy the above constraints can in principle be achieved, for example, in D-brane inflation\cite{Dvali:1998pa, Dvali:2001fw, Burgess:2001fx} where the inflaton potential has a stationary inflection point or in $\alpha$-attractor models\cite{Kallosh:2013hoa, Kallosh:2013yoa,Ferrara:2013rsa} where the inflaton potential has an exponentially flat ``wing''. However, these small-field inflation models in general suffer from initial condition problems: not only does the mean inflaton value need to sit at the flat region in the potential to initiate inflation, the initial inhomogeneity in the inflaton has to be contained around the flat region to avoid prematurely terminating inflation~\cite{East:2015ggf, Clough:2016ymm, Clough:2017efm, Aurrekoetxea:2019fhr}. 

The fact that the minimal Nelson-Barr model described in \Sec{subsec:bbp} requires small-field (and low-scale) inflation motivates study of physical mechanisms that could naturally give rise to the specific inflaton initial conditions needed to sustain inflation. On the other hand, if a tensor-to-scalar ratio greater than $ 10^{-13}$ is observed in the future and the inflation scale is proven to be greater than $10^8\,\textrm{GeV}$, minimal Nelson-Barr models will be ruled out. Further model-building is then needed to make Nelson-Barr solution to the strong CP problem compatible with higher-scale inflation. We discuss one such example of a nonminimal Nelson-Barr model that supports high-scale inflation in \Sec{sec:chiral_model}.

%%%%%%%%%%%%%%%%%%%%%%%%%%%%%%
\subsection{Implications for Topological Defects}
\label{subsec:gw}

It is possible that after inflation the universe went through other phase transitions (unrelated to CP breaking) that gave rise to new topological defects. 
Without specifying the dynamics of the phase transition (but assuming it has no impact on $\bar \theta$), we will investigate the implications of the upper bound on $\TRH$ on the gravitational wave signal produced by such defects.

We first consider a network of cosmic strings created after spontaneous breaking of some internal symmetry in the early universe. Such a network can even persist until today. 
Many experiments are proposed for probing gravitational wave signals from such a network. However, to get a detectable signal in these experiments, the string tension, which is on the same order as the temperature at which the network forms, should be $\gtrsim 10^{10}$~GeV \cite{Cui:2018rwi}.
However, the Nelson--Barr quality problem mandates much lower values for $\LCP$, which in turn suggests lower $\TRH$ values as discussed above. Thus, any detection of gravitational wave signals from a string network at these experiments would rule out minimal Nelson--Barr models.

Another possible topological defect emerging from certain symmetry breaking patterns are domain walls. We emphasize again that here we have in mind a phase transition associated with the breaking of an internal symmetry, not of CP itself, so this assumes new dynamics beyond those of the minimal Nelson--Barr model. A domain wall network energy density redshifts slower than matter, thus, unlike a string network, it will dominate the universe at some time $t_* \sim (G\sigma_{\mathrm{DW}})^{-1}$, where $\sigma_{\mathrm{DW}}$ is the energy per unit area of the domain wall. 
As a result, stable domain wall networks are already ruled out. Nonetheless, unstable domain wall networks decaying before Big Bang Nucleosynthesis (BBN) are still compatible with other observational bounds. 
This instability can be achieved via bias terms or nucleation of other defects on domain walls (for non-spacetime symmetries \cite{McNamara:2022lrw}). 
Unstable domain wall networks can still generate a detectable gravitational wave signal. The strength and the frequency range of the signal is determined by the domain walls' surface tension, $\sigma_{\mathrm{DW}}$, and their lifetime, $t_{\mathrm{dec}}$; see Refs.~\cite{Kawasaki:2011vv, Hiramatsu:2013qaa} for further details about the calculation of the signal from such networks.

\begin{figure}[t]
\centering\resizebox{0.7\columnwidth}{!}{
\includegraphics{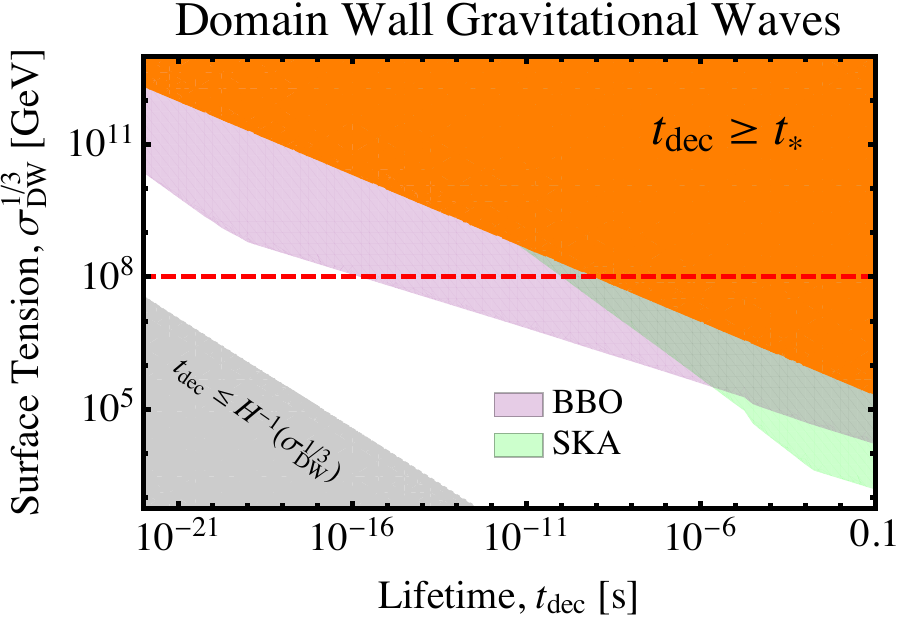}
}
\caption{The gravitational wave signal from an unstable domain wall network (unrelated to CP breaking) and the reach of various experiments in probing the parameter space. The $\TRH \lesssim 10^8$~GeV bound puts an upper bound on the surface tension $\sigma_{\mathrm{DW}}$ (red line). 
The purple (green) region can be probed by BBO (SKA), see Ref.~\cite{Moore:2014lga}. 
We find that a substantial part of the parameter space is going to be probed by future gravitational wave detectors. 
For $t_{\mathrm{dec}} \lesssim 10^{-6}$~seconds, BBO gives the best reach in the parameter space, while for higher values of $t_{\mathrm{dec}}$ SKA dominates. For every point in the orange region we have $t_{*} \leq t_{\mathrm{dec}}$, i.e., the universe enters an early domain-wall-dominant epoch, which changes the gravitational wave signal calculation. In the gray region the domain wall network lifetime is shorter than the Hubble time when the temperature of the universe is $\sigma_{\mathrm{DW}}^{1/3}$, so it promptly decays before it can give rise to a detectable gravitational wave signal.}
\label{fig:DWGW}
\end{figure}

In Fig.~\ref{fig:DWGW} we show the reach of two future gravitational wave detectors as a function of domain wall network lifetime and their surface tension (i.e., their scale of symmetry breaking). 
We consider network lifetimes $t_{\mathrm{dec}} \lesssim t_{\mathrm{BBN}} \sim 0.1$~seconds, such that the domain walls disappear before BBN, as many other bounds rule out the possibility of such long-lived defects. 
The figure clearly shows that BBO can potentially detect gravitational wave signals from domain walls with  $\sigma_{\mathrm{DW}}^{1/3} \gtrsim 10^8$~GeV (the purple region above the red line); detection of such a signal would rule out the minimal Nelson--Barr models under study. 
Nevertheless, if the quality problem could be circumvented via further model-building, i.e., $\LCP \gtrsim 10^8$~GeV was available, BBO could search a large fraction of the newly opened parameter space.

The figure also shows that, if such domain wall networks form after inflation, even with the $\sigma_{\mathrm{DW}}^{1/3} \leq \LCP \lesssim 10^8$~GeV bound, their gravitational wave signal could still be detected in a large part of the parameter space. Such gravitational wave detectors are potentially the only direct probe of physics at such high energy scales.

For large enough lifetimes, the domain walls take over the energy budget of the universe before they decay, giving rise to an early domain-wall--dominant epoch, see the orange region in Fig.~\ref{fig:DWGW}. This changes the gravitational wave signal calculation. We leave a detailed study of the reach of each experiment in this region for future works.

There have also been some studies on gravitational wave signals from hybrid defects in the early universe, e.g., see Ref.~\cite{Dunsky:2021tih} for a recent study. 
Metastable string networks that annihilate after domain walls are formed between them have less visible signals compared to a stable network. 
Nucleation of strings on domain walls can let the domain walls disappear; but this will not affect our conclusions above as they did not depend on the source of finite domain wall lifetime. 
Thus, we can safely assert that hybrid defects of strings or domain walls do not affect our previous conclusions.

%%%%%%%%%%%%%%%%%%%%%%%%%%%%%%
\subsection{Implications for Models of Baryogenesis}
\label{subsec:baryogenesis}

With $\TRH \lesssim \LCP$, the CP violation in a minimal Nelson--Barr model is effectively similar to that of the SM after inflation. 
%(We are neglecting the effect of $\theta \lesssim 10^{-10}$.) 
This can be potentially problematic for various models of baryogenesis that require sources of CP violation beyond the SM below $\TRH$.

One such model is thermal leptogenesis~\cite{Fukugita:1986hr}. 
If the SM field content is extended by heavy right-handed neutrinos,
the Nelson--Barr scenario leaves room for CP violating interactions between them and the SM. These interactions can in particular be used for generating a CP violating phase in the PMNS matrix as well. As a result, thermal leptogenesis seems to be a natural candidate for baryogenesis in Nelson--Barr setups. Below we investigate this mechanism further. 

The amount of CP breaking in decays of the right-handed neutrinos $N$ is quantified by 
\begin{equation}
\varepsilon \equiv \frac{\Gamma_{N \rightarrow H \nu}  -  \Gamma_{N \rightarrow H \bar{\nu}}}{\Gamma_{N \rightarrow H \nu}  +  \Gamma_{N \rightarrow H \bar{\nu}}},
\label{eq:defepsilon}
\end{equation}
where $\Gamma_{N \rightarrow H \nu}$ ($\Gamma_{N \rightarrow H \bar{\nu}}$) is the $N$ decay rate to the SM Higgs and neutrinos (see Ref.\cite{Covi:1996wh} for further details). 
The eventual baryon asymmetry generated in these models is directly proportional to $\varepsilon$, i.e., $\eta_\mathrm{b} \sim \varepsilon$.

Via inspection of the CP violation in this decay (assuming $\mathcal{O}(1)$ phases in the Yukawa couplings of RH neutrinos to LH neutrinos and Higgs), Ref.~\cite{Davidson:2002qv} found an upper bound on the CP breaking factor $\varepsilon$ of
\begin{equation}
\varepsilon \lesssim \frac{3}{8\pi} \frac{M_N m_\nu}{v_h^2}, 
\label{eq:DI1}
\end{equation}
where $M_N$ ($m_\nu$) is the new sterile neutrino (SM neutrino) mass. 
Given the relationship between the asymmetry $\eta_\mathrm{b}$ and $\varepsilon$ \cite{Davidson:2002qv}, this upper bound can turn into a lower bound on $M_N$. In thermal leptogenesis models, where $N$ particles start in equilibrium with the SM, we should have $\TRH \geq M_N$. 
Putting these together, and using the measured value of SM Higgs vev and SM neutrino masses, we find
\begin{equation}
\TRH \gtrsim 10^{8\,\textrm{--}\,10}~\mathrm{GeV},
\label{eq:DI2}
\end{equation}
known as the Davidson-Ibarra bound on thermal leptogenesis models.\footnote{Proposals exist for other ways of generating an asymmetry via leptogenesis that do not rely on the CP violation in the decay of the heavy neutrino, thus circumventing Davidson-Ibarra bound. See, e.g., Ref.~\cite{Domcke:2020quw} for a recent study.} 

This is in contradiction with the $\TRH \leq \LCP \lesssim 10^8$~GeV bound from results of Ref.~\cite{McNamara:2022lrw}. 
Thus, minimal Nelson--Barr models are not compatible with thermal leptogenesis proposals. 
In the upcoming section, we study a particular extension that is originally motivated by solving the quality problem of Sec.~\ref{subsec:quality_problem} but also ameliorates the tension between the thermal leptogenesis mechanism and Nelson--Barr constructions.

Another popular model of baryogenesis is electroweak baryogenesis \cite{Cohen:1990py,Cohen:1990it,Nelson:1991ab,Cohen:1993nk}. 
Electroweak baryogenesis also requires a first order electroweak phase transition, which requires new dynamics beyond minimal Nelson--Barr models as discussed in Sec.~\ref{subsec:loop_corrections}. 

Baryogenesis via the topological defects (see Refs.~\cite{Brandenberger:1994bx,Brandenberger:1994mq,Cline:1998rc} for classical examples) that originate from the spontaneous CP breaking is not viable either since these defects are inflated away. 
However, it is indeed possible to have topological defects from an unspecified symmetry breaking after inflation; these defects could give rise to gravitational waves, as discussed in the previous section, and could also play a role in generating a baryon asymmetry. 
Yet, since Nelson--Barr models are designed to constrain any new source of CP violation that interacts with the SM, it will be challenging to give rise to enough baryon asymmetry in these scenarios as well. We leave a detailed study of these mechanisms for future work.

On the other hand, models of baryogenesis where the asymmetry is generated at temperatures higher than $\TRH$, e.g., spontaneous baryogenesis \cite{Cohen:1987vi,Cohen:1988kt}, are not affected by the constraint on $\TRH$. 
It will be interesting to investigate potential relationships between the field responsible for generating a nonzero baryon number chemical potential in spontaneous baryogenesis and the $\eta_a$ fields in the Nelson--Barr setup. We will leave a detailed study of this proposal for future work.

%%%%%%%%%%%%%%%%%%%%%%%%%%%%%%
%%%%%%%%%%%%%%%%%%%%%%%%%%%%%%
\section{Chiral Nelson--Barr Models}
\label{sec:chiral_model}

As discussed in Sec.~\ref{subsec:quality_problem}, nonrenormalizable, dimension-five operators (even if Planck suppressed) can undermine the Nelson--Barr solution to the strong CP problem if $\LCP$ is too large. 
One way to avoid such dangerous operators is to assume the pair of quarks $D$, $\bar{D}$ transform {\em chirally} under a new symmetry, which we will take to be an Abelian symmetry denoted $U(1)_X$, that forbids the operators in Eq.~\eqref{eq:dangerous5}. 
This strategy was first implemented in Ref.~\cite{Valenti:2021xjp} in a model with strong dynamics. We refer to such models generally as chiral Nelson--Barr models. 

The field content for a simple realization of such a model is shown in Table~\ref{tab:chiral_charges}. Compared to the minimal model in Sec.~\ref{sec:nb_intro}, we introduce a $U(1)_X$-charged complex scalar, $\rho$, and an additional set of chiral fermions $B$ and $\bar{B}$ to cancel anomalies. The scalar $\rho$ has Yukawa couplings,
%boe%
\be
\label{eq:chiral_yukawas}
\mathcal{L} \supset -y_D \rho D \bar{D} - y_B \rho^{\dagger} B \bar{B} + \hc
\ee
%eoe%
and is assumed to acquire a vev to give masses to the chiral fermions, $\mu_D = y_D \langle \rho \rangle$. 
The $U(1)_X$ supersedes the discrete $Z_N$ symmetry required in the minimal models, but we now require at least $N_\eta \geq 2$ complex scalars, $\eta_a$, in order to generate a complex vev in the scalar potential.  For the Standard Model fields, the $U(1)_X$ charges correspond to a linear combination of hypercharge, baryon, and lepton numbers, namely $-4Y-(B-L)$.
The details of the anomaly cancellation are discussed more in Sec.~\ref{subsec:anomaly} below, and more possibilities are discussed in Appendix~\ref{app:alternatives}.

%%%%%
\begin{table}[t]
\renewcommand{\arraystretch}{1.2}
\centering
\begin{tabular}{c | c c c c }
				&  $SU(3)_c$				& $SU(2)_L$		& $U(1)_Y$ 	& $U(1)_X$ \\ \hline
$Q_i$			&  $\mathbf{3}$			& $\mathbf{2}$	& $+1/6$		& $-1$ \\
$\bar{u}_i$	&  $\mathbf{\bar{3}}$ 	& -- 					& $-2/3$		& $+3$\\
$\bar{d}_i$ 	  &$\mathbf{\bar{3}}$	& -- 					& $+1/3$		& $-1$ \\
$H$			  &--							& $\mathbf{2}$	& $+1/2$		& $-2$ \\ \hline
$D$			  &$\mathbf{3}$ 			& -- 					& $-1/3$ 		& $-1$ \\
$\bar{D}$	  &$\mathbf{\bar{3}}$	& -- 					& $+1/3$ 		& $-5$ \\
$B$			  &$\mathbf{3}$			& -- 					& $+1/3$		& $+1$ \\
$\bar{B}$	  &$\mathbf{\bar{3}}$	& -- 					& $-1/3$		& $+5$ \\
$\rho$		  &--							& -- 					& 0 				& $+6$ \\
$\eta_a$ 	  &-- 							& -- 					& 0				& $+2$ \\ \hline \hline
$\bar{e}_i$ 	  &-- 							& -- 					& $+1$			& $-5$ \\ 
$L_i$ 		  &-- 							& $\mathbf{2}$	& $-1/2$		& $3$ \\ \hline
$\bar{N}_i$ 	  &-- 							& -- 					& 0				& $-1$ \\ 
\end{tabular}
\caption{Charge assignments for the ``chiral'' Nelson--Barr model, with one choice of charges for the $U(1)_X$ symmetry. The $D$, $\bar{D}$ and $\eta_a$ fields provide the necessary structure for the Nelson--Barr mechanism, while $B$, $\bar{B}$ are introduced to make $U(1)_X$ anomaly free. The scalar $\rho$ condenses to give masses to $D$, $\bar{D}$ and $B$, $\bar{B}$, and the SM-like leptons (including singlet neutrinos) are included for generating a phase in the PMNS matrix. Similar to the vectorlike Nelson--Barr model of Sec.~\ref{sec:nb_intro}, we need multiple generations of $\eta$ fields to generate a non-zero phase in the CKM matrix. All SM fermions and $\bar{N}$ come in three generations.}
\label{tab:chiral_charges}
\end{table} 
%%%%%

Most of the analysis of how this model solves the strong CP problem carries over from the minimal model presented in Sec.~\ref{sec:nb_intro}. 
The charge assignments in Table~\ref{tab:chiral_charges} allow for the same operators as in Eq.~\eqref{eq:NBL}, but forbid any additional renormalizable couplings between $\rho$, $B$, $\bar{B}$ and the SM fermions, other than the Yukawa couplings in Eq.~\eqref{eq:chiral_yukawas}. Crucially, the quark mass matrix in Eq.~\eqref{eq:MD_tree} is unchanged, other than an additional diagonal mass term for $B$, $\bar{B}$, $\mu_B = y_B \langle \rho \rangle$, which can be taken to be real, and therefore does not affect the computation of $\thetabar$. 
Obtaining a large CKM phase requires
%boe%
\be
\label{eq:large_ckm_phase_chiralnb}
\mu_D = y_D \langle \rho \rangle \lesssim f \LCP .
\ee
%eoe%
The one-loop contributions to $\thetabar$ are also identical to those discussed in Sec.~\ref{sec:nb_intro}.

%%%%%

By construction, the dangerous dimension-five operators in Eq.~\eqref{eq:dangerous5} are forbidden by the $U(1)_X$ symmetry. The charge assignments in Table~\ref{tab:chiral_charges} also forbid any additional dimension-five operators involving $\rho$, $B$ or $\bar{B}$.
Nonrenormalizable operators that enter the fermion mass matrix 
appear only at dimension-six. These operators are
%boe%
\be
\label{eq:nonrenorm6}
\eta_a^{\dagger} \eta_b \rho D \bar{D}, \qquad
\eta_a^{\dagger} \rho\, Q_i H^c \bar{D}, \qquad
\eta_a \eta_b \eta_c^{\dagger} D \bar{d}_j .
\ee
%eoe%
The last operator in Eq.~\eqref{eq:nonrenorm6} contributes to $\mathcal{M}_{D\bar{d}}^{(1)}$, and therefore does not affect $\thetabar$, see Eq.~\eqref{eq:thetacorrection}. 
The other operators, however, do lead to contributions suppressed by a cutoff, $\Lambda_{\textrm{EFT}}^2$. 
The size of these contributions depend on the same assumptions about flavor and the size of the dimensionless parameters in the theory as discussed in the dimension-five case in Sec.~\ref{subsec:quality_problem}, but a naive estimate (assuming minimal flavor violation) yields an effect of order
%boe%
\be\label{eq:delta_theta_dim6}
\Delta \thetabar \simeq \frac{1}{y_D} \frac{\LCP^2 }{\Lambda_{\textrm{EFT}}^2} .
\ee
%eoe%
If the cutoff $\Lambda_{\textrm{EFT}}$ is taken to be the Planck scale (and assuming $y_D \sim 1$) the current bound on $\thetabar$ forces $\LCP \lesssim 10^{13}\,\textrm{GeV}$, which significantly alleviates the quality problem compared to the vectorlike, minimal Nelson--Barr constructions.

%%%%%%%%%%%%%%%%%%%%%%%%%%%%%%
\subsection{Anomaly Cancellation}
\label{subsec:anomaly}

Our strategy for relaxing the quality problem relies crucially on introducing a new symmetry that is chirally realized on the heavy quark fields. In general, this strategy results in anomalies, due to the chiral nature of $D$ and $\bar{D}$ under the new symmetry, but consistency demands that this symmetry be good up to very high scales. 
These anomalies must be removed, either by introducing additional $U(1)_X$-chiral fermions, or by appealing to the Green-Schwarz mechanism \cite{Green:1984sg}. 

In Table~\ref{tab:chiral_charges}, we have presented a specific realization of a chiral extension to the minimal Nelson--Barr model, where a single set of additional quarks, $B$ and $\bar{B}$, are sufficient to cancel all of the mixed gauge and gravitational anomalies involving $U(1)_X$ and the SM gauge group.
The key observation is that $D$, $\bar{D}$ and $B$, $\bar{B}$ form vectorlike pairs under the SM gauge group, and while the pairs are chiral under $U(1)_X$, $D$/$\bar{D}$ and $B$/$\bar{B}$ have opposite $X$ charges, so that many of the potential anomalies vanish automatically.
The charge assignments in Table~\ref{tab:chiral_charges} are a specific choice from eight separate, two-parameter families of possible assignments that cancel all of the mixed anomalies while allowing the renormalizable operators necessary for the Nelson--Barr mechanism and forbidding all of the dangerous dimension-five operators in Eq.~\eqref{eq:dangerous5}. More details on alternative charge assignments, as well as other possibilities for constructing a viable chiral Nelson--Barr model, are presented in Appendix~\ref{app:alternatives}.

%%%%%%%%%%%%%%%%%%%%%%%%%%%%%%
\subsection{Phenomenological Implications}

In Sec.~\ref{sec:cpgauge} we argued that to solve the CP domain-wall problem in Nelson--Barr setups, we should have $\TRH \lesssim \Lambda_{\mathrm{CP}}$, i.e., the epoch of inflation occurs after spontaneous CP breaking in the early universe. 
Combined with the Nelson--Barr quality problem, $\Lambda_{\mathrm{CP}} \lesssim 10^8$~GeV in minimal flavor violation constructions (or even more stringent bounds in models with nonminimal flavor), this suggests that the minimal, vectorlike Nelson--Barr models have specific implications for early universe cosmology as discussed in Sec.~\ref{sec:cosmology}. 
Our chiral setup, on the other hand, pushes this bound to $\TRH,~\Hinf \lesssim \LCP \lesssim 10^{13}$~GeV. (The  same bound was obtained in a rather different chiral model, but for parametrically the same reason, in Ref.~\cite{Valenti:2021xjp}.)
In this section we discuss how this new upper bound changes the cosmology of our chiral Nelson--Barr construction compared to the minimal, vectorlike setup of Sec.~\ref{sec:nb_intro}.

Since in our chiral setup the universe can reach higher $\TRH$, we also expect a larger tensor-to-scalar ratio $r$, see Eq.~\eqref{eq:rbound}, even up to the boundary of future experiments' reach \cite{BICEPKeck:2022mhb}. 
Larger values of $r$ are also compatible with a broader set of inflation models as surveyed in Ref~\cite{Planck:2018jri}. 
With $\Hinf \lesssim 10^{13}$~GeV, the bound on the inflaton potential flatness is relaxed to $\epsilon \lesssim 10^{-4}$ as well. 
All in all, our chiral Nelson--Barr construction is compatible with a larger set of inflation models, which can potentially give rise to detectable signals in upcoming experiments.

Furthermore, in these models, topological defects from breaking of an unspecified internal symmetry, can arise at higher energies and give rise to stronger gravitational wave signals. 
In particular, it is now possible to have string networks whose gravitational wave signal are detectable (provided the associated phase transition takes place at $T \gtrsim 10^{10}$~GeV \cite{Cui:2018rwi}). 
Detection of such signals would imply a lower bound on $\Hinf$ of inflation which would, in turn, put a lower bound on the tensor-to-scalar ratio for single-field inflation models, as well as on $\LCP$.

In Sec.~\ref{subsec:gw} we also showed that if $\TRH \gtrsim 10^8$~GeV is viable, a large parameter space of domain wall networks (from the breaking of internal symmetries, not CP) with detectable gravitational wave signals will open up, providing a well-motivated target for future detectors such as BBO, see Fig.~\ref{fig:DWGW}. 
Given the large scale of $\LCP$ required to allow for such gravitational wave signals (either from domain walls or strings), the upcoming gravitational wave detectors are the prime candidate for probing this part of our chiral model parameter space.

Finally, unlike the minimal vectorlike setup, our chiral model is now compatible with the Davidson-Ibarra bound $\TRH \gtrsim 10^{8\,\textrm{--}\,10}~\mathrm{GeV}$.  
Hence, in a chiral Nelson--Barr construction, the observed baryon asymmetry can be generated via thermal leptogenesis. 

Let us elaborate more on this mechanism. The sterile neutrinos $\bar{N}$ and $U(1)_X$ charge assignments in Table~\ref{tab:chiral_charges} allow the Lagrangian terms
\begin{equation}
\mathcal{L} \supset (\lambda_e)^i{}_j L_i H^c \bar{e}^j - (\lambda_\nu)^i{}_j L_i H \bar{N}^j - f_N^a  \eta_a \bar{N}^i \bar{N}_i.
\label{eq:Lleptons}
\end{equation}
The first term is the SM Yukawa that leads to a mass for the charged leptons; the second term is the SM Higgs coupling to SM neutrinos and the new sterile ones, while the last term generates a Majorana mass for the new sterile neutrinos after spontaneous CP breaking. 
This mass will be complex and will also break lepton number. 
Since the leptons do not interact with gluons, they do not contribute to $\bar{\theta}$ through higher loop corrections or nonrenormalizable operators.

These interactions generate a mass for the SM neutrinos via the type-I seesaw mechanism~\cite{Minkowski:1977sc}. It can also give rise to the observed baryon asymmetry of the universe via thermal leptogenesis. This requires the $\bar{N}$ particles to be in thermal equilibrium with the rest of the SM bath, which in turn suggests that their Majorana mass is below $\TRH$.
The existence of CP violation and lepton number violation (via their Majorana mass term) in the sterile neutrino decay gives rise to a net lepton number in the universe; see Sec.~\ref{subsec:baryogenesis}. 
The generated lepton asymmetry will then partially convert to a baryon asymmetry, explaining the observed $\eta_{\mathrm{b}}$ today~\cite{Workman:2022ynf}. 

Exact $\bar{N}$ masses that give rise to the observed asymmetry today depend on their abundance before they decay. 
CP violation is also expected to appear in the neutrino mass-mixing matrix, but measurements of the phase $\delta_{\textrm{PMNS}}$ from T2K and NO$\nu$A are currently in some tension~\cite{T2K:2021xwb, NOvA:2021nfi}. 
Similar to the quark sector, our setup can be used to explain any CP phase in the PMNS matrix. 
Once the dust settles on the measurement of this phase, it will be interesting to see if our setup can simultaneously accommodate neutrino masses, the observed baryon asymmetry, and this phase.

%%%%%%%%%%%%%%%%%%%%%%%%%%%%%%
%%%%%%%%%%%%%%%%%%%%%%%%%%%%%%
\section{Conclusions}
\label{sec:conclusions}

We reiterated the results of Ref.~\cite{McNamara:2022lrw} arguing that the domain walls arising from spontaneous CP breaking are stable and cannot disappear via string nucleation. 
This suggests that in such theories the reheating temperature after inflation should be below the spontaneous CP breaking scale, so that these domain walls are inflated away. We focused on the case of CP broken by fields that directly interact with SM particles (as opposed to in a hidden sector) and discussed its phenomenological implications in the early universe. 

We argued that the bound on the scale of inflation translates into an upper bound on the tensor-to-scalar ratio, suppressing it below reach of any experiment in the foreseeable future. This, in turn, rules out many possible models of inflation. 
Furthermore, we argued that detection of gravitational wave signals from any string network would rule out the minimal Nelson--Barr constructions; a substantial part of the parameter space of models that gives rise to unstable domain walls (from breaking of an internal discrete symmetry) with a detectable gravitational wave signal is also ruled out (in minimal Nelson--Barr models) thanks to the upper bound on the reheating temperature after inflation.

Nelson--Barr models, by construction, constrain the possible origin of CP breaking terms. We thus expect that in such models, the amount of CP violation after inflation is the same as predicted in the SM. This puts constraints on many well-motivated models of baryogenesis including electroweak baryogenesis or baryogenesis via defects, while models of spontaneous baryogenesis remain viable. 
We also argued that the minimal Nelson--Barr constructions with new vectorlike heavy quarks are in tension with the thermal leptogenesis mechanism, owing to the Davidson-Ibarra bound.

We also proposed a simple extension of this setup where the new fermions are chiral under a new $U(1)_X$ symmetry and showed that this setup alleviates the Nelson--Barr quality problem to $\LCP \lesssim 10^{13}$~GeV, which in turn makes it compatible with the thermal leptogenesis mechanism. 
This setup can simultaneously solve the strong CP problem, explain the observed baryon asymmetry of the universe, accommodate the nonzero neutrino masses, and explain the origin of the CP violating phases in CKM and PMNS matrices. 
Furthermore, unlike the minimal, vectorlike models, our setup is compatible with the existence of detectable string network gravitational wave signals. 

There are many clear directions for future work. For example, it would be interesting to investigate the compatibility of Nelson--Barr setups with other proposed mechanisms for baryogenesis (which generally require an additional source of CP violation) in more detail.
Furthermore, while we showed that thermal leptogenesis can be consistently embedded in a chiral Nelson--Barr model, a detailed study of its experimental signatures would be useful. 
Given the numerous interesting signals of such setups in the early universe, a more detailed study of the existing terrestrial bounds on the scale of spontaneous CP breaking is in order.

It should be emphasized that we have focused on a particularly simple model that implements the Nelson--Barr mechanism, in order to make our discussion of the quality problem---and its resolution by means of a chiral symmetry---as explicit as possible. As is clear from the discussion in \Subsec{subsec:loop_corrections}, however, this model is not natural from the standpoint of the electroweak hierarchy. We have also taken the mixed quartic $\lambda_{ab}$ to be very small (${\sim 10^{-7}}$; see discussion around~\eqref{eq:delta_theta_loop} and~\eqref{eq:lambdaloopcorrection}), which may seem to undermine the goal of explaining the small parameter $\thetabar$. However, we expect that the small quartic could naturally be explained in a supersymmetric completion, which could also ameliorate the hierarchy problem. (An alternative approach, using compositeness to suppress Yukawa couplings, was discussed in Ref.~\cite{Valenti:2021xjp}.) It would be of great interest to examine some of our conclusions in the context of a more sophisticated model that includes a mechanism for solving the electroweak hierarchy problem and suppressing the radiative corrections, along the lines of Ref.~\cite{Dine:2015jga}.

Finally, we note that Ref.~\cite{McNamara:2022lrw} pointed out that an alternative cosmic history, in which CP breaking occurs in a hidden sector before inflation, and is communicated to the SM through very weak interactions, could also solve the domain wall issue in Nelson--Barr constructions. 
In this setup CP can be spontaneously broken for a second time in the visible sector below $\TRH$, but interactions with the hidden sector give rise to an explicit source of CP violation that destabilizes the domain wall network formed at a lower scale. 
A study of the early universe cosmology in this setup, such as the compatibility with various mechanisms for baryogenesis mechanisms, or the gravitational wave signals from low-scale CP domain walls, is an interesting avenue of research.

%%%%%%%%%%%%%%%%%%%%%%%%%%%%%%
%%%%%%%%%%%%%%%%%%%%%%%%%%%%%%
\subsection*{Acknowledgments}

We are grateful to Patrick Draper and Graham Kribs for useful discussions, and to Alessandro Valenti and Luca Vecchi for useful feedback and for pointing us to their relevant work on chiral models.
PA is supported in part by the U.S. Department of Energy under Grant Number DE-SC0011640. 
The work of SH, QL, and MR is supported in part by the DOE Grant DE-SC0013607 and by the Alfred P.~Sloan Foundation Grant No.~G-2019-12504. QL and MR are also supported by the NASA Grant 80NSSC20K0506.

%%%%%%%%%%%%%%%%%%%%%%%%%%%%%%%%%%%%%%%%%%%%%%%%%%%%%%%%%%
\appendix

%%%%%%%%%%%%%%%%%%%%%%%%%%%%%%
%%%%%%%%%%%%%%%%%%%%%%%%%%%%%%
\section{Alternatives for Chiral Nelson--Barr Model Building}
\label{app:alternatives}

In this appendix we discuss alternative implementations of the chiral Nelson--Barr model presented in Table~\ref{tab:chiral_charges}, including some general comments on model building in this direction.

%%%%%%%%%%%%%%%%%%%%%%%%%%%%%%
\subsection{Alternative Charge Assignments}

We first present the most general set of charge assignments for the $U(1)_X$ symmetry for the chiral extension of the ``minimal'' BBP model described in Sections~\ref{sec:nb_intro} and \ref{sec:chiral_model}. 

Without loss of generality, we take the scalars $\eta_a$ to have $U(1)_X$ charge $+2$.
Extending the $U(1)_X$ symmetry to act on both the SM fermions (including the singlet neutrino fields, $\bar{N}$) and Higgs, the required Yukawa couplings permit us to fix the $X$ charges of all the SM fermions such that all anomalies---including the mixed gauge and gravitational anomalies---cancel exactly amongst the SM fields. This is accomplished by taking the $U(1)_X$ charges of the SM fields to be a linear combination of hypercharge $Y$ and baryon minus lepton number $B-L$.
As a result, a net contribution to the anomalies can arise only from the charges of $D$, $\bar{D}$ and $B$, $\bar{B}$. 
Up to a re-scaling, there exist several two-parameter families of charge assignments that fulfil this requirement, which we have (arbitrarily) parameterized in terms of the $U(1)_X$ charges of the Higgs and $\rho$ fields, respectively denoted by $x$ and $y$. 
There is one ambiguity arising from whether the ``Majorana''-like mass term for $\bar{N}$ comes from a Yukawa coupling to $\eta$ or $\eta^{\dagger}$, which fixes $[\bar{N}] = \pm 1$, and affects the appropriate choice of charge for $L$ and $\bar{e}$, but otherwise has no impact. Because we are gauging the product $U(1)_Y \times U(1)_X$ of two $U(1)$ symmetries, we can redefine $X$ charges by multiples of $Y$ without changing the underlying charge lattice. As a result, the parameter $x$ ultimately has no physical effect, except specifying the basis in which our gauge fields are defined.

The key to canceling the net contribution from the new fermions fields is the ``pair-wise'' vectorlike nature of the two pairs $D$, $\bar{D}$ and $B$, $\bar{B}$. Each pair of fermions separately transforms as a vectorlike pair under the SM gauge group, which immediately cancels all potential anomalies involving only the SM gauge group and gravity. 
The hypercharge of $D$, $\bar{D}$ is fixed by the renormalizable Yukawa couplings necessary for the Nelson--Barr mechanism, and the hypercharge of $B$ and $\bar{B}$ is then fixed by requiring the $U(1)_X \times U(1)_Y^2$ to cancel. This leaves a sign ambiguity on the hypercharge of $B$, $\bar{B}$.
The $U(1)_X$ charges of $D$, $\bar{D}$ and $B$, $\bar{B}$ are then fixed to be opposite in sign, so that the $U(1)_X^3$ and mixed anomalies involving only one factor of $U(1)_X$ all cancel immediately. The only remaining, nontrivial condition is the $U(1)_X^2 \times U(1)_Y$ anomaly, which fixes the difference in the charges of $D$ and $\bar{D}$ (or $B$ and $\bar{B}$) in terms of $x$.

The resulting families of solutions we have identified are summarized in Table~\ref{tab:all_chiral_charges}. 
A binary choice of generating mass for $\bar{N}$, $B$, and $D$ from either $\eta$ and $\rho$ or their conjugates gives rise to eight different families. 
Avoiding dangerous dimension-five operators, e.g., $\eta^{(\dagger)} \eta^{(\dagger)} D \bar{D}$, requires $|y| \neq 0,4$. 
The choice of $x = -2$ leads to integer values of all the charges with the smallest magnitude; we have presented the fourth column solution of Table \ref{tab:all_chiral_charges} with this choice (and with $y=6$) in Sec.~\ref{sec:chiral_model}.

%%%%%
\begin{table}[t]
\resizebox{\columnwidth}{!}{
\renewcommand{\arraystretch}{1.2}
\centering
\begin{tabular}{c | c | c | c | c | }
				& Solution 1 & Solution 2 & Solution 3 & Solution 4  \\ \hline
$Q_i$			& $\frac{1}{3}(x+1)$ & $\frac{1}{3}(x+1)$ & $\frac{1}{3}(x-1)$ & $\frac{1}{3}(x-1)$  \\
$\bar{u}_i$	& $-\frac{1}{3}(4x+1)$ & $-\frac{1}{3}(4x+1)$ & $\frac{1}{3}(-4x+1)$ & $\frac{1}{3}(-4x+1)$ \\
$\bar{d}_i$ 	& $\frac{1}{3}(2x-1)$ & $\frac{1}{3}(2x-1)$ & $\frac{1}{3}(2x+1)$ & $\frac{1}{3}(2x+1)$ \\
$H$			& $x$ & $x$ & $x$ & $x$ \\ \hline
$D$			& $-\frac{1}{3}(2x+5)$ & $-\frac{1}{3}(2x+5)$ & $-\frac{1}{3}(2x+7)$ & $-\frac{1}{3}(2x+7)$  \\
$\bar{D}$	& $\frac{1}{3}(2x+3y+5)$ & $\frac{1}{3}(2x-3y+5)$ & $\frac{1}{3}(2x+3y+7)$ & $\frac{1}{3}(2x-3y+7)$ \\
$B$			& $\frac{1}{3}(2x+5)$ & $\frac{1}{3}(2x+5)$ & $\frac{1}{3}(2x+7)$ & $\frac{1}{3}(2x+7)$  \\
$\bar{B}$	& $-\frac{1}{3}(2x+3y+5)$ & $-\frac{1}{3}(2x-3y+5)$ & $-\frac{1}{3}(2x+3y+7)$ & $-\frac{1}{3}(2x-3y+7)$ \\
$\rho$		& $y$ & $y$ & $y$ & $y$ \\
$\eta_a$ 	& 2 & 2 & 2 & 2 \\ \hline \hline
$\bar{e}_i$ 	& $2x+1$ & $2x+1$ & $2x-1$ & $2x-1$   \\
$L_i$ 		& $-x-1$ & $-x-1$ & $-x+1$ & $-x+1$  \\ \hline
$\bar{N}_i$ 	& $1$ & $1$ & $-1$ & $-1$ \\
\end{tabular}
}
\caption{$U(1)_X$ charges of different fields in our chiral model for four different anomaly-free solutions with $+1/3$ hypercharge for the $B$ field. If we set $y=-3x=6$ in the fourth column, we get the charge assignment in Table~\ref{tab:chiral_charges}. 
We can swap the ($U(1)_Y$ and $U(1)_X$) charges of $B$ and $\bar{B}$ to get four further solutions. 
After canceling all of the anomalies, we still find two degrees of freedom that parameterize all the charges in each solution; here these degrees of freedom are the Higgs charge ($x$) and $\rho$ charge ($y$). The Standard Model fields have $U(1)_X$ charge given by $2 x Y + (B-L)$ in the first two columns and $2 x Y - (B-L)$ in the last two columns. 
}
\label{tab:all_chiral_charges}
\end{table} 
%%%%%

%%%%%%%%%%%%%%%%%%%%%%%%%%%%%%
\subsection{More General Possibilities}

In Sec.~\ref{sec:chiral_model} we presented an economical extension of the ``minimal'' Nelson--Barr solution with an anomaly-free chiral symmetry, involving only two additional fermions, $B$ and $\bar{B}$. 
While it may be possible to introduce enough chiral matter to remove any anomalies in more complicated Nelson--Barr constructions, it may be challenging to do so without introducing new renormalizable (or dimension-five) operators that spoil the delicate Nelson--Barr solution, and these constructions may not be the most economical.

A massless gauge theory with anomalous matter content is simply inconsistent, but a massive gauge theory with anomalous matter content is a valid effective theory up to a UV cutoff scale that is proportional to the gauge boson mass. Such an effective theory can either be completed by introducing massive, anomaly-canceling fermions~\cite{Preskill:1990fr} (see also~\cite{Craig:2019zkf}); or via the 4d Green-Schwarz mechanism~\cite{Green:1984sg,Dine:1987xk}, in which the gauge field eats an axion $\theta$ that also has a $\theta F \widetilde{F}$ coupling. In some cases the Green-Schwarz EFT is simply a low-energy description below the scale at which charged fermions have been integrated out, but it can also be fundamental. Such theories commonly arise from compactification of higher-dimensional gauge theories with Chern-Simons terms.

Let us consider a chiral Nelson--Barr model with an anomalous $U(1)_X$ gauge symmetry that forbids all of the dimension-five operators that pose a quality problem. We assume that new physics responsible for canceling the anomaly arises at the cutoff, $\Lambda_{\textrm{EFT}}$. Focusing first on the $U(1)_X^3$ anomaly, this effective theory is consistent as long as~\cite{Preskill:1990fr}
%boe%
\be\label{eq:cutoff_bound}
\Lambda_{\textrm{EFT}} \lesssim \frac{64\pi^3}{g_X^3 \mathcal{A}_{U(1)_X^3}} m_{A'}, 
\ee
%eoe%
where $g_X$ is the gauge coupling of the $U(1)_X$ symmetry, $m_{A'} \sim g_X \langle \rho \rangle$ is the mass of the gauge boson, and $\mathcal{A}_{U(1)_X^3}$ is coefficient of the anomaly.
We see that it is difficult to push the cutoff arbitrarily far above the scale of spontaneous CP breaking to suppress the contributions from dimension-six operators unless the gauge coupling is relatively small, or if $\langle \rho \rangle \gg \LCP$, which lifts the tree-level mass for the gauge boson, relaxing the bound on the cutoff. 

A similar, more problematic constraint arises from the mixed anomalies, since the SM couplings are not free parameters. For instance, the mixed $SU(3)_c^2 \times U(1)_X$ mixed anomaly leads to a bound similar to Eq.~\eqref{eq:cutoff_bound} with two factors of $g_X$ in the denominator replaced by the strong coupling, $g_s$, and $\mathcal{A}_{U(1)_X^3}$ replaced by the mixed-anomaly coefficient, $\mathcal{A}_3$. Plugging the bound into Eq.~\eqref{eq:delta_theta_dim6}, the contribution to $\thetabar$ scales as
%boe%
\be
\Delta\thetabar \sim \frac{\alpha_s^2}{(4\pi)^4} |\mathcal{A}_3|^2 \frac{1}{y_D} \frac{\LCP^2}{\langle \rho \rangle^2} .
\ee
%eoe%
The $\alpha_s^2 / (4\pi)^4$ prefactor is $\mathcal{O}(10^{-7})$, so we need $\langle \rho \rangle \gg \LCP$ to adequately suppress these contributions. This requirement is somewhat exacerbated by the requirement in Eq.~\eqref{eq:large_ckm_phase_chiralnb}, which competes with the factor of $y_D$ in the denominator, and this suppression therefore requires some amount of tuning. The construction of UV-complete models where the chiral symmetry that mitigates the Nelson--Barr quality problem arises naturally is therefore a particularly interesting direction for future study.

The case where the 4d Green-Schwarz mechanism cancels the anomaly through a fundamental axion has mostly been studied in the context of supersymmetric theories. In these cases, the axion is part of a complex scalar field $T = \tau + \frac{i}{2\pi} \theta$, and certain forbidden dimension-five operators could be resurrected using factors of $\mathrm{e}^{-2\pi T}$ in the superpotential. Because $\langle \tau \rangle = 4\pi/g_X^2$, these are exponentially small instanton effects, and thus are expected to be safe despite arising at dimension five. In such models, a $D$-term constraint will generally enforce that some other $U(1)_X$-charged scalar gets a high-scale VEV to compensate for the VEV of the modulus. This could be the scalar $\rho$ in our scenario. Thus, dimension-five operators could naturally contribute at an exponentially small level, but at the same time the $U(1)_X$ breaking scale and hence the scale of CP breaking would be tied to UV physics near the compactification scale. Such a scenario might be of interest if one supersymmetrizes the Nelson--Barr scenario to ameliorate the electroweak hierarchy problem.

\end{spacing}

%%%%%%%%%%%%%%%%%%%%%%%%%%%%%%
%\clearpage
\addcontentsline{toc}{section}{References} 
{\small
\bibliographystyle{utphys}
\bibliography{nb-refs}
}

\end{document}